\documentstyle[12pt]{article}
\title{The Complex of  Solutions of the Nested  Bethe Ansatz. The
$A_2$ Spin Chain.}
\author{G.P. Pronko \\
{\it Institute for High Energy Physics, Protvino,}\\
{\it Moscow reg., 142284, Russia.}\\
\and 
{\it International Solvay Institute, Brussels, Belgium}\\
\and \\
Yu.G. Stroganov \\
{\it Institute for High Energy Physics, Protvino,}\\
{\it Moscow reg., 142284, Russia.}}
\date{}
\begin{document}
\maketitle

\begin{abstract}
The full set of polynomial solutions of the nested Bethe Ansatz is
constructed for the case of $A_2$ rational spin chain. The structure
and properties of these associated solutions are more various then in
the case of usual $XXX$ ($A_1$) spin chain but their role is similar.   
\end{abstract}

\begin{center}{\bf1. Introduction}
\end{center}

In our previous paper \cite{PS98} we have considered the famous
Baxter $T-Q$ equations \cite{Bax} for the simplest cases of $XXX$ and
$XXZ$ spin chains. In particular, we have shown that for each
solution of Bethe equations there exists the associated solution with
the same eigenvalue of the transfer matrix $T(\lambda)$. This
solution does not define any Bethe state, however its consideration
is proven to be  very usefull.  
  
The matter is that the associated polynomials $Q(\lambda)$ and
$P(\lambda)$ form the full set of solutions of $T-Q$ equation.
\begin{equation}
\label{baxter}
 T(\lambda) Q(\lambda)=(\lambda-i/2)^N Q(\lambda+i) +
 (\lambda+i/2)^N Q(\lambda-i),
 \end{equation}
which may be considered as the second order finite difference
equation with respect to $Q(\lambda)$.
Similar to  the case of the second order differential equations we
can express the coefficients of (\ref{baxter}) via its solutions
$Q,P$:
\begin{equation}
\label{mainold}
P(\lambda + i/2) Q(\lambda - i/2) - P(\lambda - i/2)
Q(\lambda + i/2) = \lambda^N,
\end{equation}
\begin{equation}
\label{t}
T(\lambda) = P(\lambda+i) Q(\lambda-i) - P(\lambda-i) Q(\lambda+i).
\end{equation}
   It is remarkable that the equation (\ref{mainold}) may serve as
   the
  starting point for construction of the eigenvalues of the transfer
matrix $T$. The set of polynomial solutions of this equation
reproduces the spectrum $T(\lambda)$ via (\ref{t}).

The considered construction corresponds to the case
  when the fundamental set of quantum operators belongs to the
  algebra $A_1$ (and its deformations). 
In the present paper we start the qeneralization of our approach
to the case of algebras $A_n$. For the sake of simplicity we limit
ourselves with $A_2$ case. 
Here  we consider only the case of isotropic $A_2$ spin chain leaving
the deformed case for future publications.
We shall show that each solution of the nested Bethe ansatz equations
is associated to five additional solutions corresponding to the same
eigenvalue of transfer matrix.
Also it will be shown that the  third order finite difference
equation, which is an analogue of Baxter equation for the case $A_2$
has the full set of polynomial solutions $Q,P,R$.
The corresponding "wronskian" has the  following form:
\begin{equation}
det \left|\begin{array}{ccc}
Q(\lambda-i)&Q(\lambda)&Q(\lambda+i) \\
P(\lambda-i)&P(\lambda)&P(\lambda+i) \\
R(\lambda-i)&R(\lambda)&R(\lambda+i) \end{array}
\right| =  \lambda^N .
\end{equation}
The polynomial solutions of this equation form the full spectrum of
$A_2$ transfer matrices. For example, the eigenvalues of transfer
matrices, corresponding to two fundamental representations are given
by:
\begin{equation}
\label{intTpm}
det \left|\begin{array}{ccc}
Q(\lambda-3i/2)&Q(\lambda \pm i/2)&Q(\lambda+3i/2) \\
P(\lambda-3i/2)&P(\lambda \pm i/2)&P(\lambda+3i/2) \\
R(\lambda-3i/2)&R(\lambda \pm i/2)&R(\lambda+3i/2) \end{array}
\right| = T^{\pm}(\lambda).
\end{equation}
 Apparently these equations substitute the equations (\ref{mainold})
and (\ref{t}) for the case of $A_2$.

\newpage
\begin{center}{\bf2.Various formulations of
nested Bethe Ansatz}
\end{center}

The exact formulation of the model one can find e.g. in \cite{St79}.

Diagonalization of transfer matrix and corresponding Hamiltonian has
been done with the help of so called nested Bethe Ansatz
\cite{S75}, which could be constructed in the frameworks of QISM (see
e.g.  \cite{KR81}).

Let us remind the general setup of the nested Bethe Ansatz equations
for the case $A_2$ spin chain.
Let $N$  be the length of the chain and $n_1$ and  $n_2$ -
are nonnegative integers subject to the conditions:
\begin{equation}
n_1 \le N/3, \quad n_2 \le 2 N/3, \quad 2 n_1 \le n_2.
\end{equation}
The corresponding Bethe state is defined by $n_1 + n_2$ 
parameters which we denote as:
\begin{equation}
\lambda_j^{(1)}\quad (j = 1,2, \ldots n_1),\qquad \lambda_k^{(2)}
\quad (k = 1,2, \ldots n_2).
\end{equation}
The equations for $\lambda_j^{(1)}$ and $\lambda_k^{(2)}$ have the
following form:
\begin{eqnarray}
\label{BetheA2}
&&\prod_{j^{\prime}=1}^{n_1}
\frac{\lambda_j^{(1)}-\lambda_{j^{\prime}}^{(1)}+i}
     {\lambda_j^{(1)}-\lambda_{j^{\prime}}^{(1)}-i} \times
  \prod_{k^{\prime}=1}^{n_2}
\frac{\lambda_j^{(1)}-\lambda_{k^{\prime}}^{(2)}-\frac{i}{2}}
     {\lambda_j^{(1)}-\lambda_{k^{\prime}}^{(2)}+\frac{i}{2}}=-1,
\quad (j = 1,2, \ldots n_1) \nonumber \\
&&\prod_{j^{\prime}=1}^{n_1} 
\frac{\lambda_k^{(2)}-\lambda_{j^{\prime}}^{(1)}-\frac{i}{2}}
     {\lambda_k^{(2)}-\lambda_{j^{\prime}}^{(1)}+\frac{i}{2}} \times
  \prod_{k^{\prime}=1}^{n_2}
\frac{\lambda_k^{(2)}-\lambda_{k^{\prime}}^{(2)}+i}
     {\lambda_k^{(2)}-\lambda_{k^{\prime}}^{(2)}-i} =
-\biggl(\frac{\lambda_k^{(2)}+\frac{i}{2}}
             {\lambda_k^{(2)}-\frac{i}{2}}\biggr)^N,\\
&&\quad  (k = 1,2, \ldots n_2)
\nonumber
\end{eqnarray}
Let us define the pair of polynomials $Q_1(\lambda)$ and
$Q_2(\lambda)$ of the degrees $n_1$ and  $n_2$ correspondingly by the
equations:
\begin{equation}
Q_1(\lambda) = \prod_{j^{\prime} = 1}^{n_1} (\lambda -
\lambda_{j^{\prime}}^{(1)}),\quad Q_2(\lambda) = \prod_{k^{\prime} =
1}^{n_2} (\lambda - \lambda_{k^{\prime}}^{(2)}).
\end{equation}
Making use of these polynomials we can present the equation
(\ref{BetheA2}) in the following form:
\begin{eqnarray}
\label{BetheQf}
&&Q_1(\lambda_j^{(1)}+i) Q_2(\lambda_j^{(1)}-\frac{i}{2})
 +Q_1(\lambda_j^{(1)}-i) Q_2(\lambda_j^{(1)}+\frac{i}{2}) = 0, \quad
(j = 1,2, \ldots n_1); \nonumber\\
&& (\lambda_k^{(2)}+\frac{i}{2})^N Q_1(\lambda_k^{(2)}+\frac{i}{2})
Q_2(\lambda_k^{(2)}-i) + \\
&& +(\lambda_k^{(2)}-\frac{i}{2})^N Q_1(\lambda_k^{(2)}-\frac{i}{2})
Q_2(\lambda_k^{(2)}+i) = 0.\quad  (k = 1,2, \ldots n_2).\nonumber
\end{eqnarray}
If all roots of the polynomials $Q_1(\lambda)$ and  $Q_2(\lambda)$
are simple, the equations (\ref{BetheQf}) implies:
\begin{eqnarray}
\label{nested1}
&&Q_2(\lambda+\frac{i}{2}) Q_1(\lambda-i)+Q_2(\lambda-\frac{i}{2})
Q_1(\lambda+i) = T_2(\lambda) Q_1(\lambda),\\
\label{nested2}
&&(\lambda+\frac{i}{2})^N Q_1(\lambda+\frac{i}{2}) Q_2(\lambda-i)+
  (\lambda-\frac{i}{2})^N Q_1(\lambda-\frac{i}{2})
  Q_2(\lambda+i)=\nonumber\\
&& = T_1(\lambda) Q_2(\lambda),
\end{eqnarray}
where  $T_1(\lambda)$ and  $T_2(\lambda)$ are new polynomials whose
meaning will be clear in the sequel.

The eigenvalues of transfer matrix $T(\lambda)$ enter into the game
via the folowing construction.
Shifting the argument in the equation (\ref{nested1}) by $\pm
\frac{i}{2}$ and combining the result with the equation
(\ref{nested2}), we obtain:
\begin{eqnarray}
&&\{T_1(\lambda)+(\lambda \pm \frac{i}{2})^N Q_1(\lambda \mp
\frac{3i}{2})\} Q_2(\lambda) = \\
&&\{(\lambda \pm \frac{i}{2})^N T_2(\lambda \mp \frac{i}{2})
+ (\lambda \mp \frac{i}{2})^N Q_2(\lambda \pm i)\} Q_1(\lambda \mp
\frac{i}{2}),\nonumber
\end{eqnarray}
Suppose now that $Q_2(\lambda)$ and $Q_1(\lambda \pm
\frac{i}{2})$ are mutually simple (have no common roots).
Then the last equation implies the following important formulas:
\begin{eqnarray}
\label{factor}
&&T_1(\lambda)+(\lambda \pm \frac{i}{2})^N Q_1(\lambda \mp
\frac{3i}{2}) = T^{\pm}(\lambda) Q_1(\lambda \mp \frac{i}{2}), \\
&&(\lambda \pm \frac{i}{2})^N T_2(\lambda \mp \frac{i}{2})
+ (\lambda \mp \frac{i}{2})^N Q_2(\lambda \pm i)\} = T^{\pm}(\lambda)
Q_2(\lambda),\nonumber
\end{eqnarray}
where  $T^{\pm}(\lambda)$ are the polynomials of the degree $N$,
which are the eigenvalues of the transfer matrices corresponding to
the ajoint fundamental representations of $A_2$ the auxiliary space.

Combining four equations of the system (\ref{factor}) we can find two
relations where the polynomials $T_1(\lambda)$ and
$T_2(\lambda)$ do not enter:

\begin{eqnarray}
&&(\lambda+\frac{i}{2})^N Q_1(\lambda-\frac{3i}{2})-T^{+}(\lambda)
Q_1(\lambda-\frac{i}{2})+\nonumber\\
\label{l1}
&&+T^{-}(\lambda) Q_1(\lambda+\frac{i}{2})
-(\lambda-\frac{i}{2})^N Q_1(\lambda+\frac{3i}{2})=0, \\
&& \nonumber \\
&&\lambda^N (\lambda+i)^N Q_2(\lambda-\frac{3i}{2})-(\lambda+i)^N
T^{-}(\lambda-\frac{i}{2}) Q_2(\lambda-\frac{i}{2})+\nonumber \\
\label{l2}
&&+(\lambda-i)^N
T^{+}(\lambda+\frac{i}{2}) Q_2(\lambda+\frac{i}{2})-\lambda^N
(\lambda-i)^N Q_2(\lambda+\frac{3i}{2})=0.
\end{eqnarray}
These equations, which we shall meet once more later,
we can solve with respect to $T^{\pm}(\lambda)$:
\begin{eqnarray}
\label{double}
&&T^{\pm}(\lambda \pm \frac{i}{2}) Q_1(\lambda) Q_2(\lambda \pm
\frac{i}{2}) = \lambda^N Q_1(\lambda) Q_2(\lambda \pm
\frac{3i}{2})+\nonumber\\
&&+(\lambda \pm i)^N \{Q_1(\lambda+i)
Q_2(\lambda-\frac{i}{2})+Q_1(\lambda-i) Q_2(\lambda+\frac{i}{2})\}.
\end{eqnarray}

In contradistinction with (\ref{nested1},\ref{nested2}) these
equations are homogenious with respect to  $Q_1$ and  $Q_2$ and do
not contain auxiliary polynomials $T_1$ and $T_2$.
In the same time, (\ref{double}) are equivalent to (\ref{nested1},
\ref{nested2}). To prove this statement it is sufficient to note that
r.h.s. of (\ref{double}) should devide on to $Q_1(\lambda)$, and
$Q_2(\lambda \pm \frac{i}{2}))$.
The equations (\ref{nested1}, \ref{nested2}) are the manifestations
of these requirements.

\begin{center}{\bf2. Associated solutions of the "nested" Bethe
ansatz}
\end{center}

Let us return to the system:
\begin{eqnarray}
\label{nested1p}
&&Q_2(\lambda+\frac{i}{2}) Q_1(\lambda-i)+Q_2(\lambda-\frac{i}{2})
Q_1(\lambda+i) = T_2(\lambda) Q_1(\lambda),\\
\label{nested2p}
&&(\lambda+\frac{i}{2})^N Q_1(\lambda+\frac{i}{2}) Q_2(\lambda-i)+
  (\lambda-\frac{i}{2})^N Q_1(\lambda-\frac{i}{2})
  Q_2(\lambda+i)=\nonumber\\
&& = T_1(\lambda) Q_2(\lambda),
\end{eqnarray}
In what follows we shall consider several systems of this kind, so
let
us introduce a special short notation for it:
\begin{equation}
\label{ab}
\{Q_1,Q_2;T_1,T_2\}.
\end{equation}
Each of these two equations can be viewed as a kind of $T-Q$ Baxter
equation \cite{Bax} for some inhomogenious $XXX$ - spin chain.
For example, equation (\ref{nested1p}) may be considered as $T-Q$ for
the chain of the length $n_2$ with inhomogenities defined by the
roots of the polynomial $Q_2(\lambda)$. According to the results of
our previous paper  \cite{PS98} there exists the polynomial
$P_1(\lambda)$ of the degree $n_2-n_1+1$, which together with
$Q(\lambda)$ satisfies to:
\begin{equation}
\label{p1}
Q_2(\lambda-\frac{i}{2}) P_1(\lambda+i) +
Q_2(\lambda+\frac{i}{2}) P_1(\lambda-i) = T_2(\lambda)
P_1(\lambda).
\end{equation}
In the same time, the functions  $Q_2$ and $T_2$
which play the role of coefficients in the equation (\ref{nested1p})
may be expressed in terms of two independent solutions: $Q_1$ and
$P_1$:
\begin{eqnarray}
\label{semimain1}
&&Q_2(\lambda) = P_1(\lambda+\frac{i}{2}) Q_1(\lambda-\frac{i}{2}) -
P_1(\lambda-\frac{i}{2}) Q_1(\lambda+\frac{i}{2}),\\
&&T_2(\lambda) = P_1(\lambda+i) Q_1(\lambda-i) -
P_1(\lambda-i) Q_1(\lambda+i).\nonumber
\end{eqnarray}
The second equation (\ref{nested2p}) also may be considered as $T-Q$
equation but for spin chain of the length $N+n_1$. Now the polynomial 
$\lambda^N Q_1(\lambda)$ serves as  inhomogenity. Again, according to
(\cite{PS98}) the second solution $P_2(\lambda)$ is the polynomial of
the degree $N+n_1-n_2+1$:
\begin{equation}
\label{p2}
(\lambda-\frac{i}{2})^N Q_1(\lambda-\frac{i}{2}) P_2(\lambda+i) +
(\lambda+\frac{i}{2})^N Q_1(\lambda+\frac{i}{2}) P_2(\lambda-i) =
T_1(\lambda) P_2(\lambda).
\end{equation}
The construction similar to (\ref{semimain1}) yields:
\begin{eqnarray}
\label{semimain2}
&&\lambda^N Q_1(\lambda) = P_2(\lambda+\frac{i}{2})
Q_2(\lambda-\frac{i}{2}) -
P_2(\lambda-\frac{i}{2}) Q_2(\lambda+\frac{i}{2}),\\
&&T_1(\lambda) = P_2(\lambda+i) Q_2(\lambda-i) -
P_2(\lambda-i) Q_2(\lambda+i).\nonumber
\end{eqnarray}
Let us compare the first equations from the systems
(\ref{semimain1}) and (\ref{semimain2}):
\begin{eqnarray}
\label{system}
&&Q_2(\lambda) = P_1(\lambda+\frac{i}{2}) Q_1(\lambda-\frac{i}{2}) -
P_1(\lambda-\frac{i}{2}) Q_1(\lambda+\frac{i}{2}),\\
&&\lambda^N Q_1(\lambda) = P_2(\lambda+\frac{i}{2})
Q_2(\lambda-\frac{i}{2}) -
P_2(\lambda-\frac{i}{2}) Q_2(\lambda+\frac{i}{2}),\nonumber
\end{eqnarray}
Excluding $Q_2$ from this system we obtain the factorized equation:
\begin{eqnarray}
&&Q_1(\lambda) \{ \lambda^N + P_2(\lambda-\frac{i}{2}) P_1(\lambda+i)
+ P_2(\lambda+\frac{i}{2}) P_1(\lambda-i)\} =\\
&&= P_1(\lambda) \{P_2(\lambda-\frac{i}{2}) Q_1(\lambda+i) +
P_2(\lambda+\frac{i}{2}) Q_1(\lambda-i)\}.\nonumber
\end{eqnarray}
Let $Q_1(\lambda)$ and $P_1(\lambda)$ are mutually simple (this is
equivalent to the mutual simplicity of $Q_2(\lambda)$ and
$Q_1(\lambda \pm \frac{i}{2})$). Then there exists the polynomial
$\tilde T_2(\lambda)$, satisfying to: 
\begin{eqnarray}
\label{tilde2}
&&P_2(\lambda+\frac{i}{2}) Q_1(\lambda-i)+P_2(\lambda-\frac{i}{2})
Q_1(\lambda+i) = \tilde T_2(\lambda) Q_1(\lambda),\\
&&P_2(\lambda+\frac{i}{2}) P_1(\lambda-i)+P_2(\lambda-\frac{i}{2})
P_1(\lambda+i) + \lambda^N = \tilde T_2(\lambda)
P_1(\lambda)\nonumber.
\end{eqnarray}
Remarkable that equation (\ref{p2}) and the first equation in
(\ref{tilde2}) form the new pair of equations for the nested Bethe
ansatz, which in our notations could be written as:
$\{Q_1,P_2;T_1,\tilde T_2\}$. Note that according to the first
equation of the system  (\ref{factor}), this pair corresponds to the
same eigenvalues of transfer matrices $T^{\pm}(\lambda)$, as in the
case of $\{Q_1,Q_2;T_1,T_2\}$

On the next step, starting again from the system  (\ref{system}) we
shall exclude the polynomial $Q_1$. Repeating the above procedure we
arrive at the following equations:
\begin{eqnarray}
\label{tilde1}
&&(\lambda+\frac{i}{2})^N P_1(\lambda+\frac{i}{2})
Q_2(\lambda-i)+(\lambda-\frac{i}{2})^N P_1(\lambda-\frac{i}{2})
Q_2(\lambda+i) = \nonumber \\
&& = \tilde T_1(\lambda) Q_2(\lambda),\nonumber\\
&& \\
&&(\lambda+\frac{i}{2})^N P_1(\lambda+\frac{i}{2})
P_2(\lambda-i)+(\lambda-\frac{i}{2})^N P_1(\lambda-\frac{i}{2})
P_2(\lambda+i) + \nonumber \\
&&+(\lambda-\frac{i}{2})^N (\lambda+\frac{i}{2})^N =
\tilde T_1(\lambda) P_2(\lambda)\nonumber.
\end{eqnarray}
In other words this results in the system $\{P_1,Q_2;\tilde
T_1,T_2\}$.

All discussion of this section may formulated as the following

\bf{Proposition
on the associated solutions of the nested Bethe ansatz equations.}
\rm

Let we have the solution $\{Q_1,Q_2;T_1,T_2\}$
 of the Bethe equations (\ref{nested1p},\ref{nested2p}).
 Degrees of the polynomials in the brackets are 
$(n_1,n_2;N+n_1,n_2)$ correspondingly.

Then, there exist the pair of associated solutions
$\{Q_1,P_2;T_1,\tilde T_2\}$ and $\{P_1,Q_2;\tilde T_1,T_2\}$,
for which the degrees are:
$(n_1,N+n_1-n_2+1;N+n_1,N+n_1-n_2+1)$ and
$(n_2-n_1+1,n_2;N+n_2-n_1+1,n_2)$.

\begin{center}{\bf3. Complex of the solutions of the nested Bethe
ansatz equations}
\end{center}

Each of  two associated solutions $\{Q_1,P_2;T_1,\tilde T_2\}$ and
$\{P_1,Q_2;\tilde T_1,T_2\}$,
could be considered as a result of two operations
${\cal F}_1$ ¨ ${\cal F}_2$ on the initial solution
$\{Q_1,Q_2;T_1,T_2\}$:
\begin{eqnarray}\
\label{oper}
&&{\cal F}_1 \{Q_1,Q_2;T_1,T_2\} = \{P_1,Q_2;\tilde
T_1,T_2\},\\
&&{\cal F}_2 \{Q_1,Q_2;T_1,T_2\} = \{Q_1,P_2;T_1,\tilde
T_2\}\nonumber
\end{eqnarray}

The smart reader can get an impression that there exist an infinite
set of the associated solutions. However, these two operations
${\cal F}_1$ and ${\cal F}_2$  form a finite group.
This, in turn, guarantees a finiteness of the number of the
associated
solutions.

Indeed, first of all, 
let us remark that two successive applications of  the operation
${\cal F}_1$ (${\cal F}_2$) return us back to the initial solution.
In other words these operations are the involutions:
\begin{equation}
 {\cal F}_1^2 = {\cal F}_2^2 = I.
\end{equation}
The next nontrivial operations are the products:
$ {\cal F}_2 {\cal F}_1 $, $ {\cal F}_1 {\cal F}_2 {\cal F}_1$
et cetera and the set with replacement $ {\cal F}_1 \Leftrightarrow
{\cal F}_2 $.
The key statement is that this set is actually finite because $ {\cal
F}_1$ and $ {\cal F}_2$ satisfy to Artin relation:

\begin{equation}
\label{Artin}
{\cal F}_1 {\cal F}_2 {\cal F}_1={\cal F}_2 {\cal F}_1 {\cal F}_2
\end{equation}
This relation results in the following structure of the whole set of
nested Bethe ansatz solutions:
\begin{eqnarray}
\label{Sol}
&\{Q_1,Q_2;T_1,T_2\}& \nonumber \\
{\cal F}_1 \swarrow&&{\cal F}_2 \searrow \!\!\!\! \searrow \nonumber
\\
\{P_1,Q_2;\tilde T_1,T_2\}&&\{Q_1,P_2;T_1,\tilde T_2\} \nonumber \\
\Downarrow {\cal F}_2 &&{\cal F}_2 \downarrow \\
\{P_1,R_2;\tilde T_1,T_2^{\prime}\}&&\{R_1,P_2;T_1^\prime,\tilde
T_2\} \nonumber \\
{\cal F}_1 \searrow&&{\cal F}_2 \swarrow \!\!\!\! \swarrow \nonumber
\\&\{R_1,R_2;T_1^{\prime},T_2^{\prime}\}& \nonumber 
\end{eqnarray}
To prove this statement, first of all, recall that $T^{\pm}(\lambda)$
defined in (\ref{factor}) are invariants under ${\cal F}_1, {\cal
F}_2$ operations. 
The matter is that each of these operations do not change one of two
pairs $Q_i,T_i$ and due to (\ref{factor}) it is sufficient for
conservation of $T^{\pm}(\lambda)$. 

Now let us consider equations (\ref{l1},\ref{l2}) from the first
section: 
\begin{eqnarray}
&&(\lambda+\frac{i}{2})^N Q_1(\lambda-\frac{3i}{2})-T^{+}(\lambda)
Q_1(\lambda-\frac{i}{2})+\nonumber\\
\label{lp1}
&&+T^{-}(\lambda) Q_1(\lambda+\frac{i}{2})
-(\lambda-\frac{i}{2})^N Q_1(\lambda+\frac{3i}{2})=0, \\
&& \nonumber \\
&&\lambda^N (\lambda+i)^N Q_2(\lambda-\frac{3i}{2})-(\lambda+i)^N
T^{-}(\lambda-\frac{i}{2}) Q_2(\lambda-\frac{i}{2})+\nonumber \\
\label{lp2}
&&+(\lambda-i)^N
T^{+}(\lambda+\frac{i}{2}) Q_2(\lambda+\frac{i}{2})-\lambda^N
(\lambda-i)^N Q_2(\lambda+\frac{3i}{2})=0.
\end{eqnarray}
These equations may be considered as linear homogenious finite
difference equations of the third order for polynomials
$Q_1$ and $Q_2$. The invariants $T^{\pm}(\lambda)$ play the role of
the coefficients.
The equations have three lineary independent solutions:
$Q_1,P_1,R_1,$ and $Q_2,P_2,R_2,$ correspondingly.
If the relation (\ref{Artin}) is not valid, i.e. the chain of
solutions (\ref{Sol}) is longer then we should obtain more then three
solutions to each equations (\ref{lp1},\ref{lp2}), what is
impossible.

\begin{center}
{\bf4. Concluding remarks}
\end{center}

In our previous paper (\cite{PS98}) we have considered two
fundamental polynomial solutions to Baxter $T-Q$-equation. These
solutions may be considered as the fundamental objects in the
integrable $A_1$ spin chain models. They give rise to all possible
fusion relations for the transfer matrices corresponding to different
spin in the auxiliary space  and the transfer matrices themselfs
could be expressed in terms of these polynomial solutions.

For the case $A_2$ spin chain we expect that six polynomial solutions
(\ref{Sol}) play the same role.
Indeed, let us remind that the polynomials $Q_1$,$P_1$ and $R_1$ are
the solutions of the equation (\ref{lp1}):
\begin{equation}
\label{linPR}
\left|\begin{array}{cccc}
 Q_1(\lambda -\frac{3i}{2})& Q_1(\lambda -\frac{i}{2})&
 Q_1(\lambda +\frac{i}{2}) & Q_1(\lambda +\frac{3i}{2}) \\
 P_1(\lambda -\frac{3i}{2})& P_1(\lambda -\frac{i}{2})&
 P_1(\lambda +\frac{i}{2}) & P_1(\lambda +\frac{3i}{2}) \\
 R_1(\lambda -\frac{3i}{2})& R_1(\lambda -\frac{i}{2})&
 R_1(\lambda +\frac{i}{2}) & R_1(\lambda +\frac{3i}{2}) 
\end{array} \right|
\left|\begin{array}{c}
(\lambda + \frac{i}{2})^N \\
- T^+(\lambda) \\ 
 T^-(\lambda) \\
- (\lambda -\frac{i}{2})^N 
\end{array} \right| = 0.
\end{equation}

Excluding $T^{\pm}(\lambda)$ from this system we get the following
equation:
\begin{equation}
\label{MAIN}
det \left|\begin{array}{ccc}
Q_1(\lambda-i)&Q_1(\lambda)&Q_1(\lambda+i) \\
P_1(\lambda-i)&P_1(\lambda)&P_1(\lambda+i) \\
R_1(\lambda-i)&R_1(\lambda)&R_1(\lambda+i) \end{array}
\right| =  \lambda^N .
\end{equation}
This equation is analog of the fundamental "wronskian" (16) from
paper (\cite{PS98}). As in the case of $A_1$ the (\ref{MAIN}) can be
considered as the starting point for construction of the polynomials
$Q_1,P_1,R_1$ and consequently the transfer matrices
$T^{\pm}(\lambda)$:
\begin{equation}
\label{Tpm}
det \left|\begin{array}{ccc}
Q_1(\lambda-3i/2)&Q_1(\lambda \pm i/2)&Q_1(\lambda+3i/2) \\
P_1(\lambda-3i/2)&P_1(\lambda\pm i/2)&P_1(\lambda+3i/2) \\
R_1(\lambda-3i/2)&R_1(\lambda\pm i/2)&R_1(\lambda+3i/2) \end{array}
\right| = T^{\pm}(\lambda).
\end{equation}

Consider for example the case of 3-sites chain $N=3$.
The full set of polynomial solutions of the equation (\ref{MAIN}) in
this case is:

 \vspace{0.5cm}

\begin{tabular}{|c|c|c|c|}
\hline
Number & $Q(\lambda)$  & $P(\lambda)$ & $R(\lambda)$ \\
\hline
1 & 1 & $\lambda$ & $\lambda^5 + \frac{5}{3} \lambda^3$  \\
\hline
2 &1 &$ \lambda^2 + \frac{\lambda}{\sqrt 3}$ & $\lambda^4 - \frac{2
\lambda^3}{\sqrt 3} - \sqrt 3 \lambda$ \\
\hline
3 &1 &$ \lambda^2 - \frac{\lambda}{\sqrt 3}$ & $\lambda^4 + \frac{2
\lambda^3}{\sqrt 3} + \sqrt 3 \lambda$ \\
\hline
4 & $\lambda$ & $\lambda^2 +  \frac{1}{3}$ & $\lambda^3 $\\
\hline
\end{tabular}

\vspace{0.5cm}

These four solutions correspond to four irreducible representations
which enter into decomposition of the product of $N=3$ fundamental
representations:
\begin{equation}
{\bf 3} \times {\bf 3} \times {\bf 3} = {\bf 1} + {\bf 8} + {\bf 8} +
{\bf 10}
\end{equation}
Note that we can express in terms of these polynomials not only 
$T^{\pm}(\lambda)$ which correspond to the fundamental
representations
in the auxiliary space  but also  transfer matrices for any other
representations of $A_2$.  

Similar relations exist also for the polynomials $Q_2,P_2,R_2$.
Taking into account the first equation (\ref{semimain1}) one can 
establish that these relations are not independent.

In the paper (\cite {KLWZ97}) the authors also have considered the
$A_n$ case of nested Bethe anzats equations using the analogues of
$T-Q$ Baxter equations. However, in their approach the
"regularization" by means of "external magnetic field" is essential
and it is not known how to remove this regularization. Therefore, we
at the present time unable  to compare our results.  

{\bf Acknowledgements}		
\noindent
We are grateful to A.V. Razumov, M.V. Saveliev, S.M.
Sergeev  for useful discussion.

The research was supported in part by RFFR grant 98-01-00070 and
INTAS 96-690.

\end{document}